# Twist-bend heliconical chiral nematic liquid crystal phase of an achiral rigid bent-core mesogen

Dong Chen[a], Michi Nakata[a†], Renfan Shao[a], Michael R. Tuchband[a], Min Shuai[a], Ute Baumeister[b], Wolfgang Weissflog[b], David M. Walba[c], Matthew A. Glaser[a], Joseph E. Maclennan[a] and Noel A. Clark[a*]

[a]Department of Physics and Liquid Crystal Materials Research Center, University of Colorado, Boulder, CO 80309-0390, USA

[b]Institut fűr Physikalische Chemie, Martin-Luther-Universität Halle-Wittenberg, Műhlporte 1, D-06108 Halle (Saale), Germany.

[c]Department of Chemistry and Biochemistry and Liquid Crystal Materials Research Center, University of Colorado, Boulder, CO 80309-0215, USA.

[†]deceased

*E-mail: noel.clark@colorado.edu

**Abstract**

The chiral, heliconical (twist-bend) nematic ground state is reported in an achiral, rigid, bent-core mesogen (UD68). Similar to the nematic twist-bend ($N_{TB}$) phase observed in bent molecular dimers, the $N_{TB}$ phase of UD68 forms macroscopic, smectic-like focal-conic textures and exhibits nanoscale, periodic modulation with no associated modulation of the electron density, i.e., without a detectable lamellar x-ray reflection peak. The $N_{TB}$ helical pitch is $p_{TB}$ ~ 14 nm. When an electric field is applied normal to the helix axis, a weak electroclinic effect is observed, revealing 50 µm-scale left- and right-handed domains in a chiral conglomerate.

**Introduction**

In the entire history of liquid crystals, only four distinct nematic ground states have been found: the uniaxial, the biaxial, and, for chiral molecules, the helical nematic and blue phases. Recently, the twist-bend nematic ($N_{TB}$) phase, a fundamentally new type of nematic ground state of achiral molecules, exhibiting layer-free, helical liquid crystal ordering of nanoscale pitch, has been structurally identified and characterized [1,2]. The twist-bend nematic phase was initially suggested as a theoretical possibility by Meyer over 40 years ago [3] and discussed in 2000 by Dozov [4] and Memmer [5], who suggested that the tendency for local bend curvature in the director field of bent molecules (for example, Figure 1a) could stabilize a twisted and bent director distribution with the molecules precessing on a cone (Figure 1c). Observations of an unusual phase of bent molecular dimers, such as CB7CB (Figure 1b), in which a pair of rigid monomers are linked together by means of a flexible odd-carbon number aliphatic spacer [6,7] having nematic characteristics, spontaneous chirality [8,9], and smectic textural features [10,11] but no smectic-like lamellar x-ray diffraction peaks [12] have been interpreted in terms of the $N_{TB}$ structure [4]. The $N_{TB}$ phase is in general polar and structurally chiral despite being formed from achiral molecules. The helix periodicity has recently been directly observed in CB7CB and other dimers using freeze-fracture transmission electron microscopy (FFTEM) [1,2] which revealed that the TB helix of CB7CB, for example, has an amazingly short pitch ($p_{TB}$ ~ 8 nm), making it the liquid crystal coherent ordering that is closest in scale to the molecular size (about 3 nm in the CB7CB case).

This structure agrees with the predictions of atomistic computer simulation [1].

Here we report the observation of a twist-bend nematic phase in the rigid-bent-core liquid crystal UD68 (Figure 1a), an extraordinary new addition to the family of liquid crystals showing the $N_{TB}$ phase, and just the sort of system treated theoretically by Dozov [4] and in simulation by Memmer [5]. We provide direct evidence of the chiral, periodic, conical helical structure of UD68, using freeze-fracture transmission electron microscopy (FFTEM), x-ray diffraction (XRD), depolarized transmission light microscopy (DTLM), and electro-optical measurements. UD68 was synthesized by Schröder et al. [13], who proposed in the bent-core series in Figure 1a for n < 7 an unusual nematic-to-nematic phase transition, with the molecules in the lower temperature phase stacked in the bend direction, forming bundles of non-defined length as the precursor of the columnar phase. We show that the structural characteristics of this low temperature nematic phase are nearly identical to those found in the $N_{TB}$ phases of the bent molecular dimers.

**Experiments and Results**

The chemical structure and phase sequence of UD68 are shown in Figure 1a. When UD68 is filled in a nylon-coated cell weakly rubbed for planar alignment, the nematic comes in from the isotropic on cooling with a typical Schlieren texture with s=1 and s=1/2 defects, and visible fluctuations of the director field (Figure 2a). On further cooling, the director fluctuations disappear and the twist-bend nematic phase grows in, forming typical focal conic domains with dark brushes parallel to the polarizers (Figure 2b, top). The initially smooth, fan-like texture subsequently breaks into a cross-hatched texture (Figure 2c, right). At even lower temperature, dendritic aggregates typical of the columnar phase grow in (Figure 2c, left). In homeotropic cells, the $N_{TB}$ phase is essentially dark under crossed polarizers, suggesting that the phase is uniaxial. The nematic phase shows no x-ray layer reflection peak, but a diffuse wide-angle peak corresponding to the intermolecular spacing (Figure 2d). When cooled to the $N_{TB}$ phase, the x-ray diffraction pattern is essentially the same as that in the nematic phase, with no layer reflection peak. Although the $N_{TB}$ phase shows focal-conic textures characteristic of soft systems with one-dimensional periodic ordering, the absence of lamellar x-ray reflections indicates that there is no modulation of the electron density associated with this periodic structure. When cooled to the columnar phase, the XRD pattern shows two peaks, one at q~0.14 Å$^{-1}$ and one at q~0.15 Å$^{-1}$, corresponding to a rectangular columnar structure with lattice parameters a~4.2 nm and b~4.5 nm [13].

In a strongly rubbed, planar cell, the nematic phase is well aligned (Figure 3a), with uniform birefringence. The texture of the $N_{TB}$ phase, in contrast, is not optically uniform, growing in with stripes (Figure 3b) similar to those observed in [14]. Under an applied electric field, domains with two different molecular tilts are apparent, with distinct boundaries (Figure 3c), that are visible even at zero field (Figure 3e), similar to the domain boundaries seen in bent molecular dimers (Figure 3f), as first reported by Panov et. al [9]. When the E field is reversed, the molecular tilts in the two sets of domains are reversed and their colors interchange (Figure 3d). This chiral response is direct evidence that the $N_{TB}$ phase is a conglomerate of left- and right-handed domains. In the twist-bend configuration, the molecular dipoles of the bent-core mesogens are oriented perpendicular to the helix axis. In an applied electric field, the uniform heliconical structure becomes distorted and electroclinic tilts of opposite sign are induced in the left- and right-handed domains. The electroclinic effect in nematics has been observed previously in cholesteric liquid crystals [15]. Recently, Dozov et. al. further clarify the electroclinic effect in the twist-bend nematic phase of the bent molecular dimer CB7CB [16] which is similar to that observed in the bent-core molecules UD68.

Freeze-fracture transmission electron microscopy, a technique which enables direct visualization of the microstructure of liquid crystals on the nanometer scale [17], was used to explore further the internal structure of the $N_{TB}$ and Col phases. The one-dimensional, periodic structure corresponding to the conical helix of the $N_{TB}$ phase is clearly visible in the FFTEM images in Figures 4a and b. In contrast to typical smectic samples, it is hard to find layer surfaces in the $N_{TB}$ phase, as the molecules are intercalated in the conical helical structure and there are no distinct layer interfaces as in smectics. This observation is consistent with the fact that we are unable to draw freely-suspended film of UD68 in the $N_{TB}$ phase. The periodicity of the bent-core $N_{TB}$ phase is about p~14 nm, a little longer than the pitch measured in the bent molecular dimers CB7CB (p~8 nm) [1]. Taking the molecular length as L=44.5 Å (estimated by CPK models assuming a core bend angle of 120° and an all-trans conformation of the alkyl chains), there are only a few molecules in each helical turn (Figure 1c). At lower temperature, the N$_{TB}$ phase transitions to the Col phase. The two-dimensional periodic structure of the rectangular columnar phase is shown in Figures 4c and d, with the layer and in-plane layer modulation parallel to the fractured surface.

**Discussion**

Although dimers with odd-numbered spacers such as CB7CB superficially resemble bent-core molecules like UD68, their conformational energy landscape is very different. In the dimers, the energy difference between trans (180°) and gauche (±60°) dihedral orientations is of the order of only 0.5 kcal/mol, which at 97°C is less than 1 $k_B$T of energy, so that the alkyl spacer is quite flexible and plenty of gauche conformations are expected as a result of thermal fluctuations [1]. The CB7CB N$_{TB}$ phase therefore can easily be super-cooled to room temperature and a glass transition occurs at around 7°C [18]. The UD68 N$_{TB}$ phase with a rigid-bent-core, on the other hand, transitions to the Col phase on cooling.

So far, the helical structure have been observed in the N*, SmC*, N$_{TB}$ and SmCP phases, which are compared in detail in Table S.1 (Supporting Information) [19]. In the N* and SmC* phases, the molecules are rod-like and chiral, while the N$_{TB}$ and SmCP phases can be formed by bent, achiral molecules. Due to their rigid, bent molecular shape, the flexoelectric response resulting from linear coupling between the polar order normal to the director and bend director deformations is much larger for bent-core liquid crystals than for rod-like liquid crystals [20] and this flexoelectricity enhances the heliconical structure with twist and bend molecular deformation in the $N_{TB}$ phase [21,22]. However, in order to form the $N_{TB}$ phase instead of the SmCP, the tendency for layering must be weak or absent. One way to tune this is by altering the alkyl tail, with molecules with longer alkyl tails being more likely to form smectic layers. The homologues of the unsymmetrical bent-core molecule UD68 that have long alkyl tails (n>7) form the SmCP phase, while molecules with short alkyl tails (n<7) show the $N_{TB}$ phase. Analogous behavior is seen in dimers. For example, modification of CB7CB with alkoxy chain terminal, only dimers with short terminals (n=4) are able to form the $N_{TB}$ phase [23]. If the bent shape is lost altogether, for example in odd-numbered trimers, no $N_{TB}$ phase is observed [24].

**Summary**

The mesophases of a bent-core liquid crystal UD68 have been characterized using DTLM, XRD, FFTEM and electro-optical measurements. Due to the natural tendency of banana-shaped molecules to induce a local bend of the molecular director, this material forms the $N_{TB}$ phase with a conical twist-bend helix, a phase observed previously only in bent molecular dimers linked by flexible, odd-numbered alkyl chain. Although the liquid crystal molecules are achiral, the $N_{TB}$ phase is polar and chiral. Electroclinic effects of opposite sign reveal conglomerate domains of opposite handedness in cells. Unlike in the dimers where the $N_{TB}$ phase is easily supercooled, UD68 has an $N_{TB}$-Col phase sequence where the molecules are locked into a structure with two-dimensional order at lower temperature. Study of molecular architecture and ordering reveals that only homologues with short alkyl tails favor the twist-bend director field and nematic order, while longer homologues form conventional SmCP phases.

**Acknowledgement**

The authors thank V. P. Panov and J. K. Vij for the image in Figure 3f. This work was supported by NSF MRSEC Grant DMR-0820579.

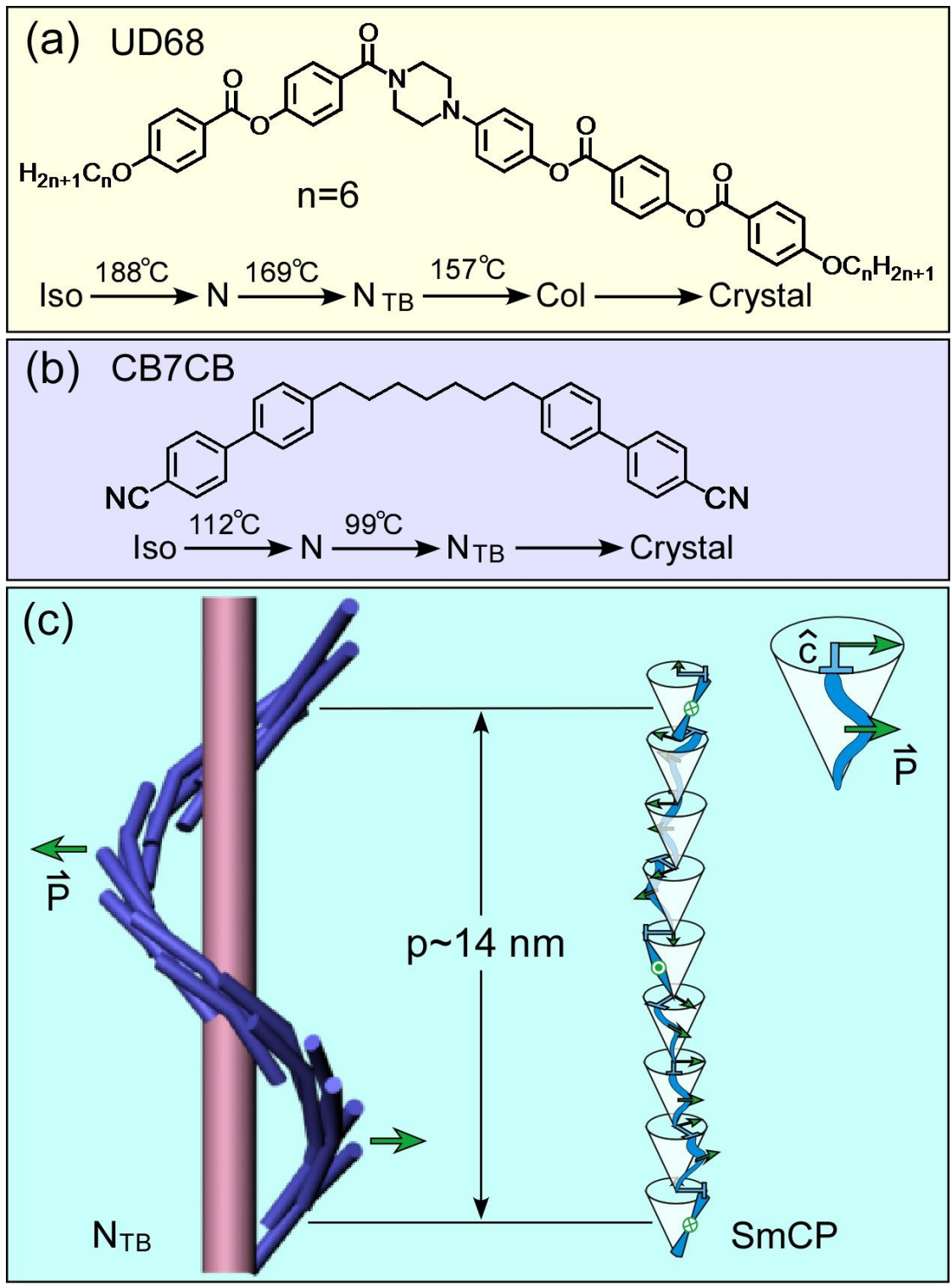

Figure 1 (color online): Chemical structures and phase sequences of the bent-core molecule UD68 (a) and the bent molecular dimer CB7CB (b). (c) Molecular organization of the $N_{TB}$ and SmCP phases of bent-core molecules. Both the $N_{TB}$ and SmCP phases are chiral and polar with the molecules forming a heliconical structure.

In the $N_{TB}$ phase, the molecules are intercalated, while in the SmCP phase the bent-core molecules form well-defined smectic layers, the combination of macroscopic polarization and molecular tilt making the layers chiral.

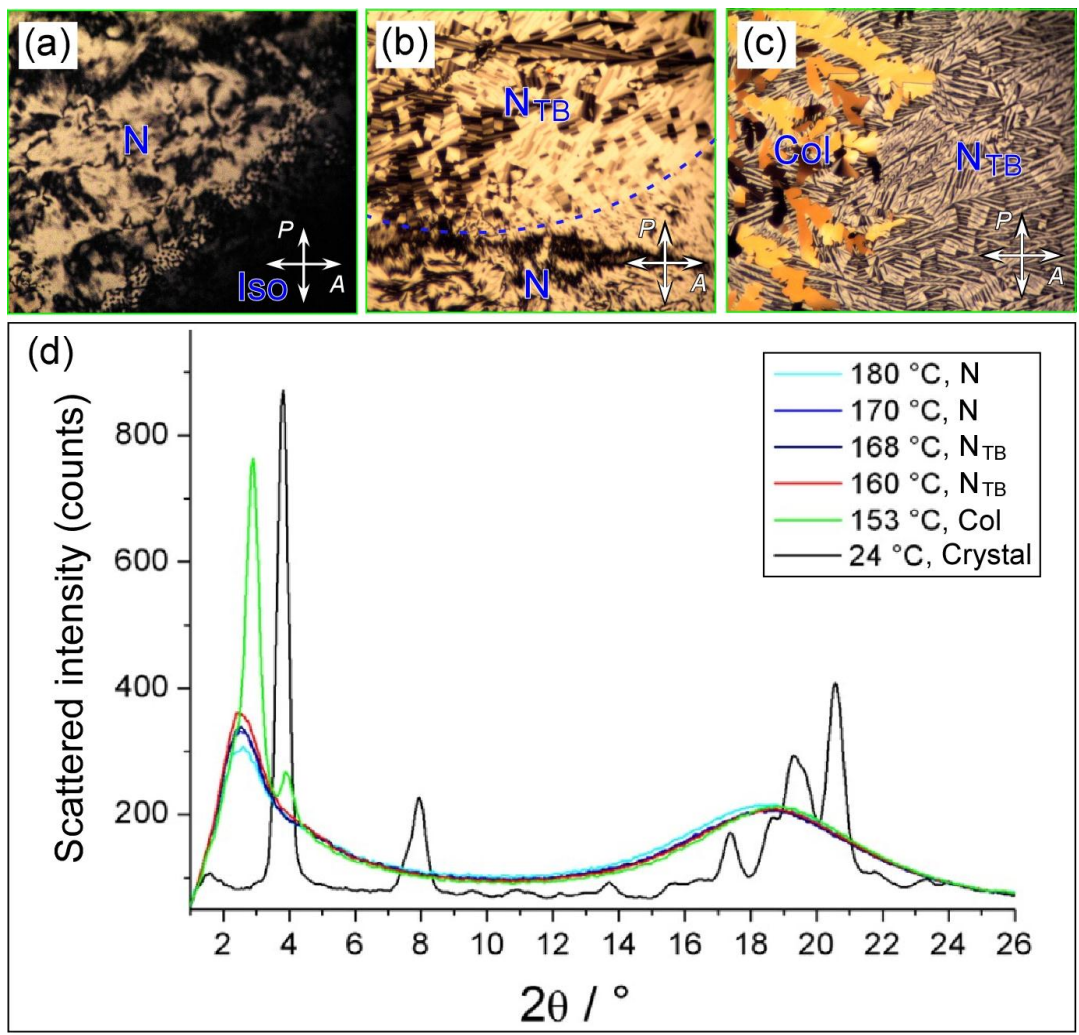

Figure 2 (color online): DTLM textures in a weakly-rubbed cell and XRD of the mesophases of UD68. (a) Nematic phase (top left) with characteristic Schlieren texture indicating random planar alignment grows in from the isotropic phase (bottom right), which is dark under crossed polarizers (T=186°C). (b) In the $N_{TB}$ mesophase (T=167°C), focal conic domains (top half) form from the nematic phase (bottom half), with the dark brushes parallel to the polarizers. (c) At lower temperature (T=155°C), yellow dendritic aggregates of the Col phase (left) grow into the $N_{TB}$ texture (right). Images (a)-(c) were obtained in a 2 μm nylon cell with weak rubbing. (d) XRD of the N, $N_{TB}$, Col, and Crystal phases. The measurements were taken on cooling from the nematic phase at 180°C, except for the measurement at 24°C, which was taken after the samples was kept at room temperature for five days.

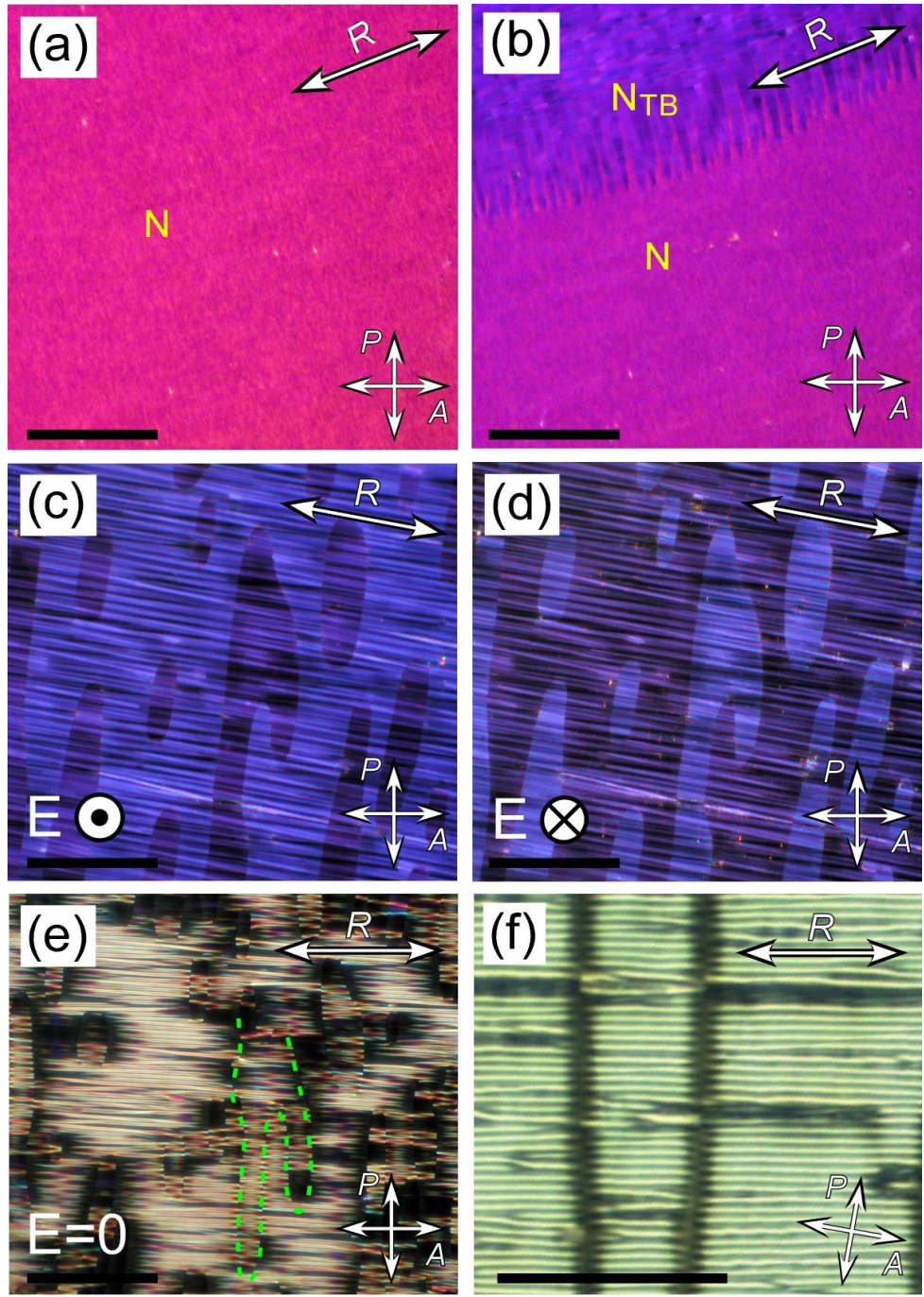

Figure 3 (color online): Microscopic textures and electro-optic response of the $N_{TB}$ phase in a strongly rubbed cell. (a) Uniformly aligned nematic phase at T=170°C with the director along the rubbing direction R. (b) Stripes are a characteristic feature of the $N_{TB}$ phase as it grows in at T=167°C. (c) In an applied E field of 20 V/μm, two regions of different molecular tilt can be distinguished, with the dark and bright regions interchanging when the E field is reversed (d). (e) In the absence of external electric field, the boundaries of domains of opposite chirality are characterized by discontinuities in the stripe pattern, for example along the dashed green line. (f) The texture of the $N_{TB}$ phase in a 2 μm planar cell of a bent molecular dimer diluted with its monomer (35 wt%) reported in [9] shows chiral domain boundaries similar to those in (e). Reproduced with permission from Applied Physics Letters **101**, 234106 (2012). Copyright 2012 American Institute of Physics. Images (a)-(e) were taken in an Instec 4 μm planar cell. The scale bars are 100 μm.

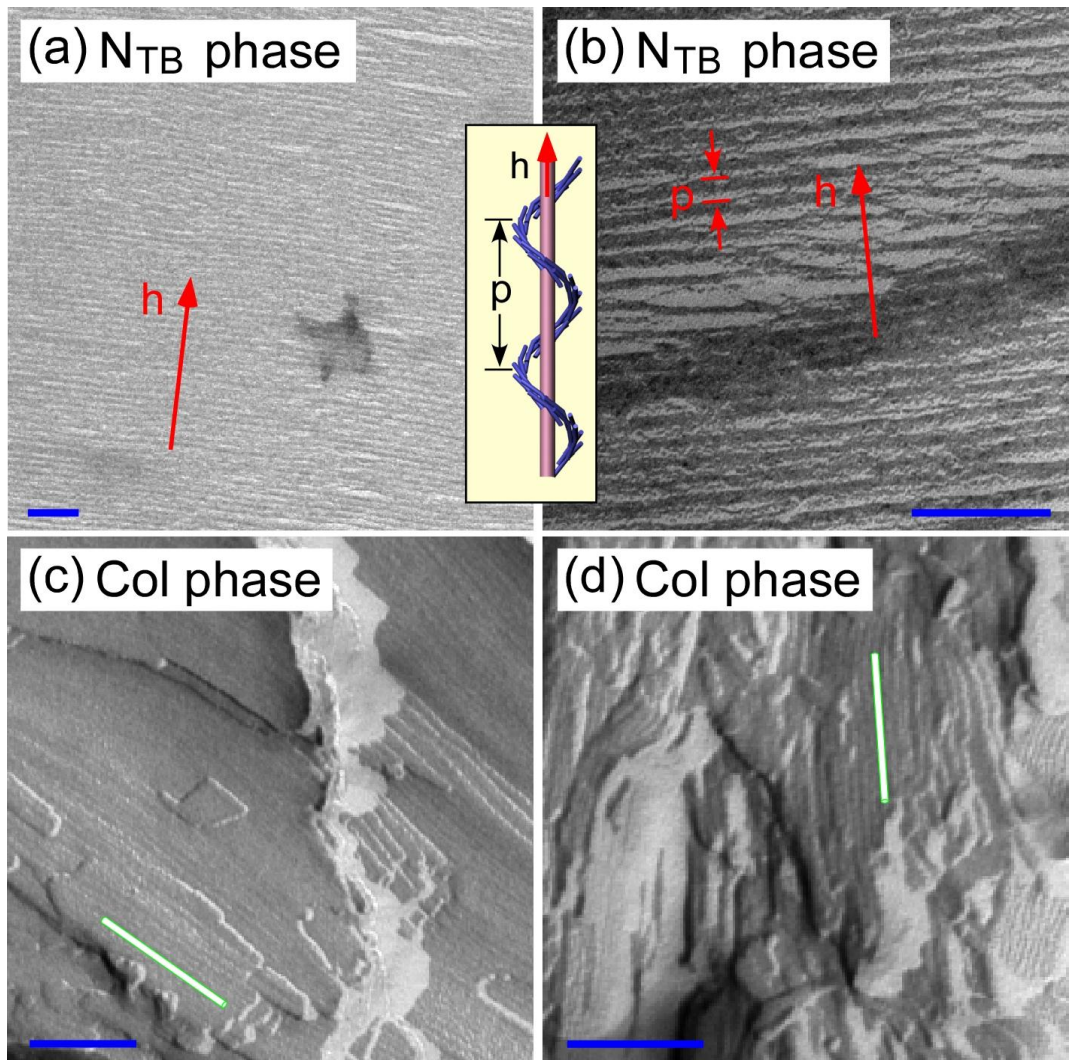

Figure 4 (color online): Freeze-fracture images of the $N_{TB}$ and Col mesophases. (a) and (b) FFTEM images of the $N_{TB}$ phase, showing the conical helical structure of period $p_{TB}$~14 nm. In both images, the helix axis h is approximately parallel to the fracture plane as modeled in the inset and indicated by the red arrows. No layer surfaces are observed in the $N_{TB}$ phase. The sample was sandwiched between two clean glass planchettes and quenched at T=167°C. (c) and (d) FFTEM images of the Col phase, quenched at T=157°C. The Col phase shows a 2D periodic structure, with layers parallel to the fracture plane and layer modulations along the layer surface. The column orientation is sketched in the inset. Distinct layer steps can be observed in the Col phase. The scale bars are all 100 μm.

# Supporting Information for Twist-bend heliconical chiral nematic liquid crystal phase of an achiral rigid bent-core mesogen

Dong Chen[a], Michi Nakata[a,†], Renfan Shao[a], Michael R. Tuchband[a], Min Shuai[a], Ute Baumeister[b], Wolfgang Weissflog[b], David M. Walba[c], Matthew A. Glaser[a], Joseph E. Maclennan[a] and Noel A. Clark[a,*]

[a]Department of Physics and Liquid Crystal Materials Research Center, University of Colorado, Boulder, CO 80309-0390, USA

[b]Institut für Physikalische Chemie, Martin-Luther-Universität Halle-Wittenberg, Mühlporte 1, D-06108 Halle (Saale), Germany.

[c]Department of Chemistry and Biochemistry and Liquid Crystal Materials Research Center, University of Colorado, Boulder, CO 80309-0215, USA.

[†]deceased

*E-mail: noel.clark@colorado.edu

Table S.1: Comparison of the N*, SmC*, $N_{TB}$ and SmCP phases, all of which form the helical structure.

|  | **N*** | **SmC*** | **$N_{TB}$** | **SmCP** |
|---|---|---|---|---|
| Molecular shape | Rod-like | Rod-like | Bent-core | Bent-core |
| Molecular chirality | Chiral | Chiral | Achiral | Achiral |
| Layers? | No | Yes | No | Yes |
| Polarization? | No | Yes | Yes | Yes |
| Twist deformation? | Yes | Yes | Yes | Yes |
| Bend deformation? | No | Yes | Yes | Yes |